\ifx\onecol\undefined
\documentclass[journal]{IEEEtran}
\else 
\documentclass[onecolumn,draftcls,12pt]{IEEEtran}
\fi

\usepackage[utf8]{inputenc}
\usepackage{physics}
\usepackage{amsmath}
\usepackage{amssymb}
\usepackage{subfigure}
\usepackage{graphicx}
\usepackage{graphics}
\usepackage{color}
\usepackage{psfrag}
\usepackage{cite}
\usepackage{balance}
\usepackage{algorithm}
\usepackage{accents}
\usepackage{amsthm}
\usepackage{bm}
\usepackage{url}
\usepackage{algorithmic}
\usepackage[english]{babel}
\usepackage{multirow}
\usepackage{enumerate}
\usepackage{cases}
\usepackage{stfloats}
\usepackage{dsfont}
\usepackage{soul}
\usepackage{amsfonts}
\usepackage{fancyhdr}
\usepackage{hhline}
\usepackage{array}
\usepackage{booktabs}
\usepackage{tabularx}
\usepackage{framed}
\usepackage{float}
\usepackage{soul}
\usepackage{url}
\usepackage{hyperref}
\usepackage{xspace}

\newcommand{\zw}[1]{{#1}}

\newcommand{\rev}[1]{{#1}}
\newcommand{\sai}{SpaceAI-Infra\xspace}

\begin{document}
\bstctlcite{CompactReferenceAuthors}

\title{Space AI Infrastructure: \\ Where the Edge Meets the Orbital Cloud}

\author{{Zhanwei Wang, Jianhao Huang, Min Sheng,~\IEEEmembership{Fellow,~IEEE,} and Kaibin Huang,~\IEEEmembership{Fellow,~IEEE}}\vspace{-5mm}
\thanks{Z.~Wang, J.~Huang, and K.~Huang are with the Department of
Electrical and Computer Engineering, The University of Hong Kong (HKU),
Hong Kong SAR, China (e-mail: zhanwei.wang@hku.hk;
Jianhaoh@hku.hk; huangkb@eee.hku.hk).
Corresponding author: K.~Huang.}\thanks{M.~Sheng is with the State Key Laboratory of Integrated Service
Networks, Xidian University (XDU),
Xi'an, Shaanxi, China (e-mail: msheng@mail.xidian.edu.cn).}}

\maketitle

\begin{abstract}

The fusion of artificial intelligence (AI) and space technologies is extending spaceborne intelligence beyond onboard payloads toward a global infrastructure, here termed Space AI infrastructure (SpaceAI-Infra).
Its promise rests on two distinctive advantages of space systems: abundant solar energy in orbit and the global reach of satellite constellations.
Realizing this promise hinges on a fundamental systems question: How should satellite constellations be designed and orchestrated to deliver data-center-scale AI services competitive with their terrestrial counterparts?
This article develops a roadmap-oriented answer.
We first chart the evolution of SpaceAI-Infra through three progressive stages: onboard AI, Space Edge AI, and Space Cloud AI.
By revisiting the space-constrained fundamentals for end-to-end token services, we derive enabling design principles and translate them into an integrated space edge--cloud architecture.
The communication-centric space edge comprises free-flying LEO constellations that provide user-proximate computing and form the access and backhaul fabric for cloud AI services.
Complementing the edge, the computing-centric space cloud comprises formation-flying satellite constellations organized as space data centers in dawn--dusk sun-synchronous orbits.
A case study demonstrates that the proposed architecture can support data-center-scale foundation-model inference for millions of concurrent users worldwide.

\end{abstract}


\section{Introduction}

\emph{Space AI infrastructure} (\sai) is emerging as an increasingly credible platform for powering energy-intensive \emph{large language model} (LLM) services. Its key feature lies in converting solar energy harvested in orbit into AI-generated tokens and delivering them to users worldwide. 
Such infrastructure will not only alleviate the growing pressure on terrestrial power systems but also extend AI services to underserved regions where deploying terrestrial data centers is economically or geographically impractical. 
Against this backdrop, leading companies in the space and AI sectors, such as SpaceX, Google, and NVIDIA~\cite{Bleier2026SpaceDatacenters}, have moved quickly to secure an early foothold in this emerging field.

\begin{figure*}[t]
    \centering
    \includegraphics[width=0.73\linewidth]{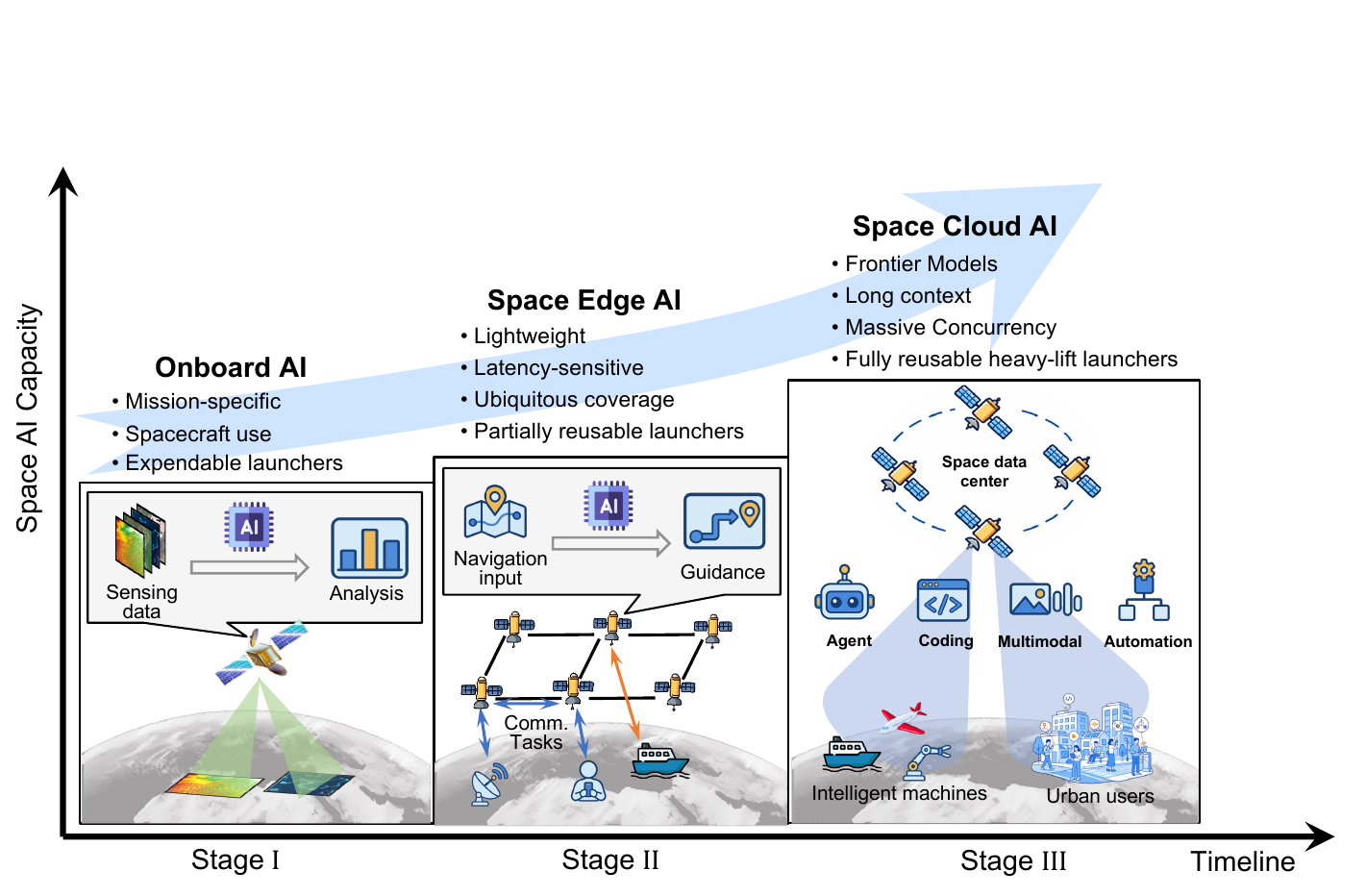}
    \caption{Roadmap of Space AI Infrastructure.
    }
    \label{fig:roadmap}
    \vspace{-5mm}
\end{figure*}

Unlike its terrestrial counterpart, \sai is following a reverse evolutionary path under aerospace and economic constraints. 
Terrestrial AI infrastructure has first matured around centralized cloud data centers and is now extending intelligence toward the network edge, closer to where data are generated~\cite{GX-CM-2020}. 
In space, by contrast, data centers with comparable capabilities face substantially greater technical and economic barriers to deployment.
At a representative mass-to-power ratio of $40~\mathrm{kg/kW}$, competing economically with terrestrial infrastructure would limit the spacecraft-build budget to only \$250--\$1000/kg, whereas the latest published rocket launch price is roughly an order of magnitude higher~\cite{Turyshev2026ODC}. This substantial disparity makes an immediate leap to data-center-scale orbital computing impractical; we therefore envision \sai evolving toward this long-term goal through the following three progressive stages, also illustrated in Fig. \ref{fig:roadmap}.

\subsubsection{Onboard AI}

The first stage is \emph{onboard AI}, in which mission-specific AI capabilities are embedded in individual \zw{satellites} for local inference. This stage addresses the downlink bottleneck arising from the mismatch between rapidly growing sensing-data volumes and the limited capacity of space--ground links. For example, a single high-resolution Earth observation satellite can generate $1$--$2~\mathrm{TB}$ of imagery per day, whereas its RF downlink operates at only $220~\mathrm{Mbit/s}$ during brief ground passes~\cite{Bleier2026SpaceDatacenters}. Even assuming uninterrupted transmission and neglecting protocol overhead, downlinking the entire volume would require approximately $10$--$20$ hours per day, far exceeding the contact time available during typical ground passes. By processing sensor data at the source, onboard AI can discard low-quality observations and extract actionable information before downlink.
This reduces communication overhead and response latency.
Consequently, onboard AI \zw{can enhance mission autonomy} through a single-satellite upgrade without requiring constellation-scale coordination.
Such single-spacecraft deployments can be supported by expendable launchers.

\subsubsection{Space Edge AI}

The second stage extends AI service provisioning from onboard tasks to \zw{users worldwide, including those on land, aboard aircraft and vessels}. It embeds computing payloads (e.g., AI functions) into communication-centric satellite constellations \zw{in \emph{low Earth orbit}} (LEO), with user-proximate connectivity, giving rise to \emph{Space Edge AI}. 
Its advantages are twofold. First, computing payloads can reuse the spacecraft bus, communication links, and operational infrastructure established for communication missions~\cite{Turyshev2026ODC}. Second, the proximity and global coverage of LEO constellations make Space Edge AI well suited to ubiquitous and latency-sensitive computing services, such as real-time translation, interactive gaming, and autonomous navigation.
Partially reusable launchers support this stage by reducing the cost of repeated launches for constellation deployment and replenishment~\cite{NASA2026LaunchServicesRockets}.

\subsubsection{Space Cloud AI}
\label{sec:space_cloud_AI}

The third stage is \emph{Space Cloud AI}, which targets general-purpose AI services through \zw{\emph{space data centers} (SpaceDCs)}. It is envisioned to support frontier foundation models, multimodal workloads, and agentic services for millions of concurrent users worldwide. Current initiatives illustrate this vision at different levels of maturity. For example, SpaceX has proposed an AI satellite capable of supplying 120~kW to its computing payload, comparable to the power demand of an NVIDIA GPU rack.
Google's Project Suncatcher aims to build a scalable distributed SpaceDC comprising TPU-equipped satellites flying in formation in a dawn--dusk \emph{sun-synchronous orbit} (SSO).
By networking multiple such satellites through high-speed optical links, Space Cloud AI could pool computing resources in orbit and ultimately provide data-center-scale AI services to terrestrial users.
Fully reusable heavy-lift launchers are envisioned to enable economical deployment of the computing, power, and thermal infrastructure required at this scale~\cite{AgueraYArcas2026Suncatcher}.

This article is a humble attempt to conceptualize the roadmap and enabling architecture for \sai.
The rest of the article is organized as follows.
Sec.~\ref{sec:fund-design} examines the orbital fundamentals underlying roadmap stages and formulates the corresponding design principles.
Building on these principles, Sec.~\ref{sec:architecture} presents \zw{an integrated space edge--cloud architecture} for \emph{end-to-end} (E2E) token services.
Sec.~\ref{sec:case_study} evaluates the resulting architecture through a case study.
Finally, Sec.~\ref{sec:conclusion} concludes the article and outlines future research directions.

\section{Fundamentals and Design Principles for Space AI Infrastructure}
\label{sec:fund-design}
In this section, we examine the physical fundamentals underlying \sai, spanning individual-satellite orbital characteristics and multi-satellite cooperation, and derive the corresponding design principles.

\subsection{Solar Harvesting and Token Delivery}
\label{sec:energy-latency}

Space-based token services face two fundamental bottlenecks: energy-intensive token generation and latency-sensitive token delivery. The former converts harvested solar energy into tokens through onboard computing, whereas the latter delivers them to requesting users. Both processes are closely coupled with the orbital configurations of token-serving satellites, which are presented as follows. 

\subsubsection{Harvesting-latency Tradeoff}
The energy harvesting efficiency is determined by the sunlit ratio during the lifetime and governed by the Sun--Earth--satellite geometry. 
One of the key parameters is orbital altitude, defined as the satellite's height above a specified Earth reference surface. For otherwise comparable orbits, increasing altitude generally \rev{reduces the fraction of each orbit spent in eclipse, thereby increasing} the annual sunlit fraction, as shown in Fig.~\ref{fig:sun-latency-tradeoff}. This energy gain comes at the cost of longer space--ground propagation. 
For an overhead satellite, moving from a $600$-km LEO to the $35{,}786$-km GEO increases the round-trip propagation delay from approximately $4$ to $240$~ms. Orbital altitude therefore governs a fundamental harvesting--latency tradeoff. Low-latency E2E token services require a computing orbit that combines abundant sunlight with short propagation paths. We next identify such a token-service-friendly orbit.

\subsubsection{Sun Synchronous Orbit}

\begin{figure}[t]
    \centering
        \subfigure[Solar availability and one-way propagation-latency tradeoff.]{
        \includegraphics[width=0.4\textwidth]{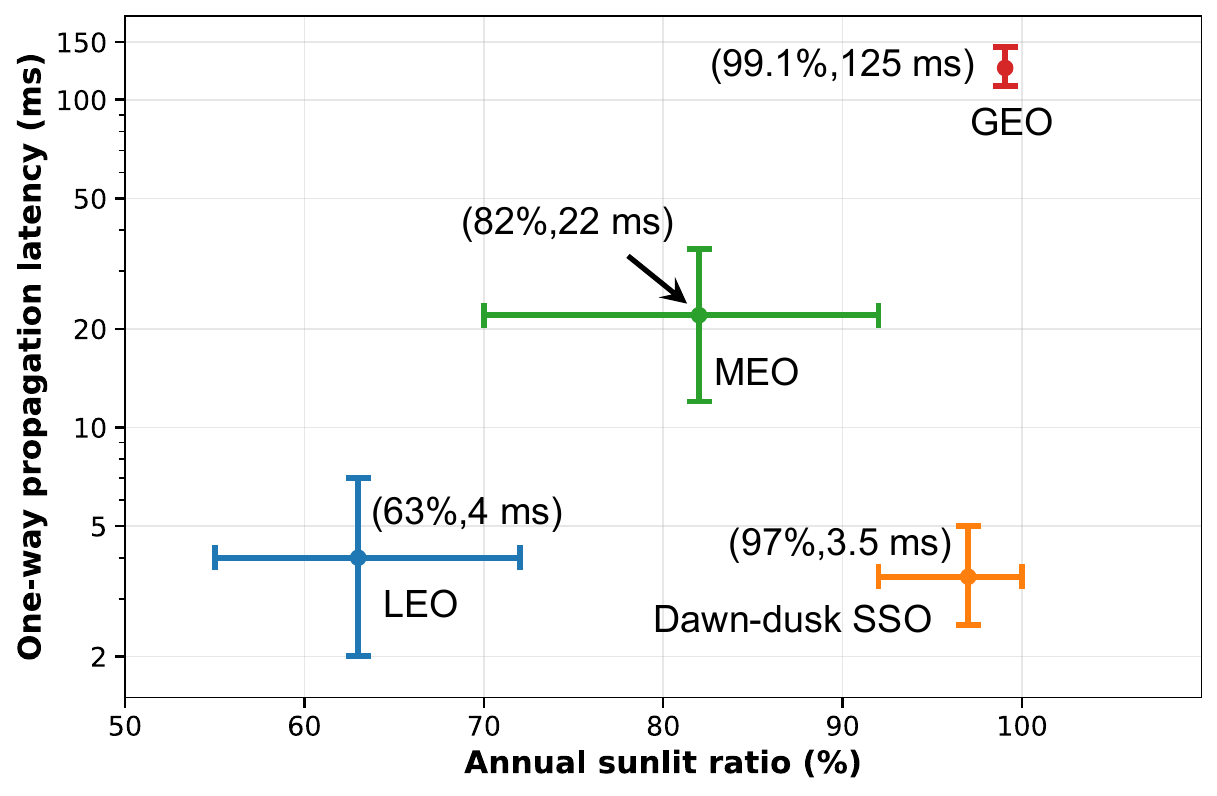}
        \label{fig:sun-latency-tradeoff}
    }
    \hfill
    \subfigure[Sun-synchronous orbit geometry.]{
        \includegraphics[width=0.3\textwidth]
        {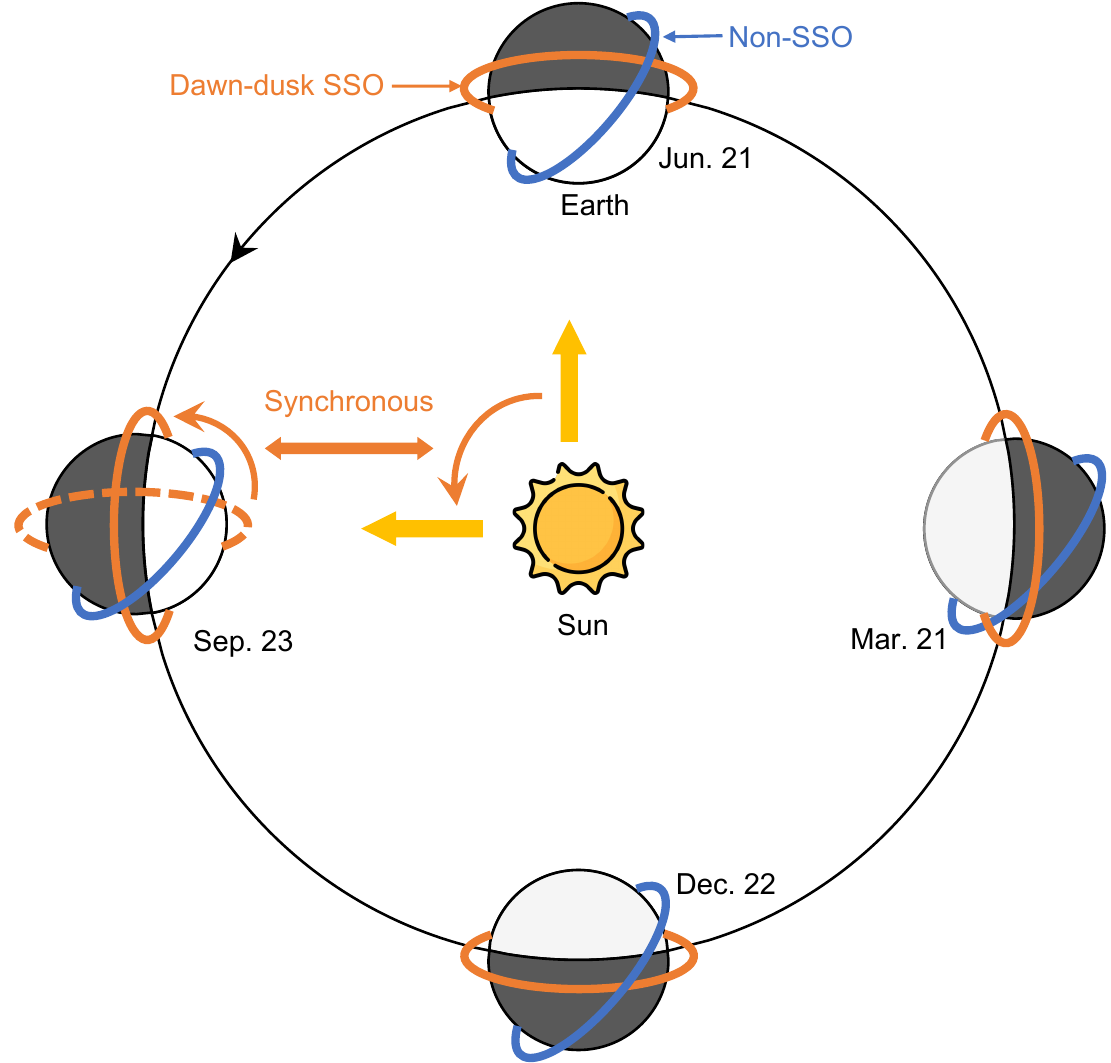}
        \label{fig:sso_demo}
    }
    
    \caption{Orbital characteristics relevant to space AI infrastructure.
    LEO spans altitudes from 200 to 2,000~km, MEO from 2,000 to
    35,786~km, and GEO is located at an altitude of 35,786~km.}
    \label{fig:orbital-characteristics}
    \vspace{-5mm}
\end{figure}

Another factor affecting sunlight availability is the solar beta angle, defined as the angle between the Sun vector and the orbital plane. 
Owing to Earth's oblateness, the orbital plane undergoes nodal precession about Earth's rotational axis.
By jointly selecting the orbital altitude and inclination, the precession rate of the orbital plane can be matched to the apparent annual motion of the Sun, approximately $360^\circ$ per year, thereby maintaining \rev{an approximately fixed local solar time at the ascending node}~\cite{Montenbruck2000SatelliteOrbits}. Such an orbit is termed \zw{an} SSO. As illustrated in Fig.~\ref{fig:sso_demo}, the Earth--Sun direction rotates throughout the year, while the orbital plane precesses at approximately the same angular rate.

Among SSO designs, dawn--dusk SSO is particularly attractive for space computing, as it mitigates the harvesting--latency dilemma. 
Its orbital plane remains close to the day--night terminator, \rev{providing large solar beta angles and near-continuous sunlight}~\cite{Turyshev2026ODC}. Moreover, this configuration can be engineered in LEO  at typical altitudes of $700$--$1{,}200$~km. 
The resulting minimum ground--satellite round-trip propagation delay is approximately $4.7$--$8.0$~ms, \rev{which is of the same order of magnitude as the single-digit-millisecond network latency reported for the AWS Los Angeles Local Zone}~\cite{Roberts2020LALocalZone}. 
This observation leads to the following orbital design principle.

{\setlength{\FrameRule}{0.6pt}\setlength{\FrameSep}{6pt}\begin{framed}
\noindent \textbf{Principle 1:}
SpaceDCs should operate in dawn--dusk SSO to combine near-continuous solar exposure with LEO-class propagation latency.
\end{framed}
}

\begin{figure*}[t]
    \centering
\includegraphics[width=0.65\linewidth]{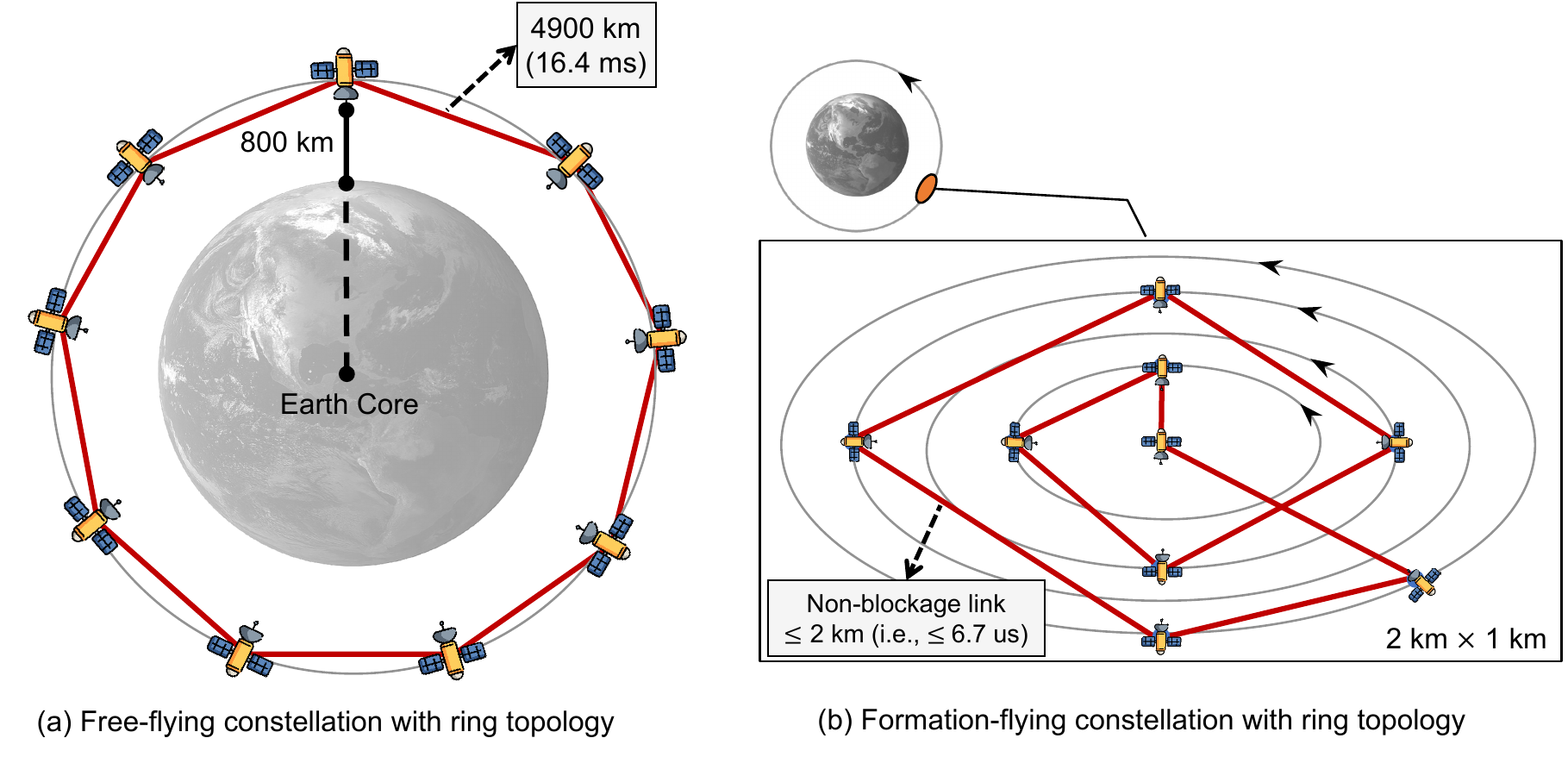}
    \caption {An illustrative comparison of free-flying and formation-flying satellite constellations, each with
a ring topology.}
    \label{fig:free-formation-comparison}
    \vspace{-5mm}
\end{figure*}

\vspace{-3mm}
\subsection{Satellite Constellations for Space AI Infrastructure}

Space-based token services at scale require capabilities far beyond what an individual satellite can support, necessitating multi-satellite cooperation.
The spatiotemporal organization of cooperating satellites is characterized by their constellation, defined by their relative positions and motion over time. The formation type governs the inter-satellite communication and cooperative computing that underpin token services. We next distinguish two representative formation types suited to Space Edge AI and SpaceDCs, respectively.

\subsubsection{LEO-based Free-flying Constellations}
A \textit{free-flying constellation} comprises spatially distributed satellites that independently follow prescribed orbits without maintaining a fixed relative geometry. Such constellations are well suited to communication-centric missions and offer three advantages for space-based token services:
\begin{itemize}
  \item \textit{Ubiquitous coverage:}
Distributing satellites across multiple orbital planes, as in Walker, rosette, and polar constellations, enables persistent connectivity over wide geographical areas. By optimizing orbital-plane placement and satellite phasing, LEO-based free-flying formation allows minimizing the number of satellites required to achieve a target coverage level, motivating their adoption in commercial systems such as Iridium NEXT, OneWeb, and Starlink.

    \item \textit{Low-latency computing access:}
    As discussed in Sec. \ref{sec:energy-latency},
    deployment in LEO shortens space--ground propagation paths, allowing terrestrial users to access onboard computing with low latency. For example, a satellite at the approximately $550$-km altitude of a Starlink shell incurs a minimum ground--satellite round-trip propagation delay of around $4$~ms.

    \item \textit{Token relay for SpaceDCs:}
    LEO constellations can serve as an access and relay layer between terrestrial users and the higher-altitude dawn--dusk SSO data centers identified above. Their wide-area coverage and inter-satellite connectivity allow user requests to be forwarded to SpaceDCs and the generated tokens to be returned.
\end{itemize}
These properties align with the vision of Space Edge AI, leading to the following design principle.

{\setlength{\FrameRule}{0.6pt}\setlength{\FrameSep}{6pt}\begin{framed}
\noindent \textbf{Principle 2:}
LEO-based free-flying constellations are well suited to Space Edge AI and should form the network backbone of \sai.
\end{framed}
}

However, this coverage-oriented geometry is less suitable for tightly coupled cooperative computing. As illustrated in Fig.~\ref{fig:free-formation-comparison}(a), satellites in a free-flying constellation are typically uniformly distributed along an orbital plane to maximize coverage.
Consider a ring topology between them with each satellite maintaining two intra-plane inter-satellite links~\cite{three-body-constellation-IOTJ}. 
At an orbital altitude of $800$~km, adjacent satellites are separated by approximately $4{,}900$~km. When onboard memory constraints require an LLM to be partitioned across multiple satellites, each decoded token must traverse at least one inter-satellite link as hidden states pass between model partitions. This transfer alone adds at least $16.4$~ms to the per-token latency, thereby limiting token-generation throughput.

\subsubsection{Formation-flying Constellations}

Unlike its free-flying counterpart, a \textit{formation-flying constellation} maintains multiple satellites in a prescribed relative geometry. Their relative motion is coordinated through orbital control. Formation flying offers two principal advantages:
\begin{itemize}
    \item \textit{Proximity-enabled computing:}
    The compact geometry substantially shortens inter-satellite distances and hence reduces propagation latency for cooperative computing. Figure~\ref{fig:free-formation-comparison}(b) illustrates a representative design based on Google's Project Suncatcher~\cite{AgueraYArcas2026Suncatcher}. The formation comprises a chief satellite near the center of a $2\times1$-km region and multiple deputy satellites following controlled elliptical trajectories relative to the chief. A maximum inter-satellite separation of $2$~km incurs only approximately $6.7~\mu$s of propagation delay, comparable to that of terrestrial data-center interconnects.

    \item \textit{Modular scalability and resilience:}
Reconfigurable ISLs allow satellites to be added to, removed from, or isolated within the computing cluster. This flexibility supports incremental capacity expansion and technology refresh by deploying advanced satellites and retiring obsolete ones. When satellites become unavailable because of debris impacts or space-weather disturbances, workloads can be rerouted among the remaining nodes, while only the affected satellites need to be replaced. Formation flying thus improves service continuity and reduces recovery costs compared with a monolithic orbital platform~\cite{Bleier2026SpaceDatacenters}.
\end{itemize}

As envisioned \zw{for the third stage in Sec.~I}, Space Cloud AI must support frontier-scale models, complex AI workloads, and massive user concurrency. Their aggregate computing and memory demands typically exceed the capacity of a single satellite. Agentic AI services, for example, invoke numerous tools and specialized models, while multi-agent orchestration requires coordinated execution and intermediate data exchange across multiple agents~\cite{Chaudhry2026Murakkab}. Mapping such workloads onto \sai therefore requires multiple satellites interconnected through high-bandwidth, low-latency links. Formation-flying constellations are well suited to these requirements, leading to the following design principle.

{\setlength{\FrameRule}{0.6pt}\setlength{\FrameSep}{6pt}\begin{framed}
\noindent \textbf{Principle 3:}
Formation-flying constellations should form the computing backbone of SpaceDCs, enabling tightly integrated distributed AI services in \sai.
\end{framed}
}

\subsection{Why the Space Edge Must Meet the Orbital Cloud}

The preceding principles reveal a fundamental complementarity between Space Edge AI and Space Cloud AI. LEO-based free-flying constellations provide ubiquitous access and low-latency token delivery, but their communication-centric power budgets constrain onboard computing scale, particularly during orbital eclipses. Formation-flying SpaceDCs can instead pool computing and memory resources under near-continuous sunlight, yet their compact, compute-centric geometry does not provide ubiquitous user coverage. Neither infrastructure alone can satisfy the combined coverage and computing requirements of Stage III. Realizing this stage therefore requires an edge--cloud-integrated \sai that couples the coverage-oriented space edge with the computing-oriented orbital cloud, as developed in the next section.

\section{Space Edge-cloud Architecture}
\label{sec:architecture}

In this section, building on the preceding design principles, we present an edge--cloud-integrated architecture for \sai.

\subsection{Architecture Overview}

As shown in Fig.~\ref{fig:sdc-constellation}, \sai comprises two subnetworks with complementary roles.

\subsubsection{Space Data Center Subnetwork}

The SpaceDC subnetwork comprises multiple SpaceDCs deployed in dawn--dusk SSOs, each implemented as a formation-flying satellite constellation. Benefiting from near-continuous solar exposure, each satellite is envisioned as a high-capacity computing node comparable to a GPU rack equipped with multiple accelerators. Interconnected through high-speed ISLs, these satellites collectively provide data-center-scale computing capacity.
Within each SpaceDC, satellites are functionally classified as gateway satellites and general-purpose computing satellites. The former serve as both external interfaces and internal orchestrators. They maintain optical links with the space edge subnetwork to receive AI requests and return generated tokens. 
Internally, they schedule cooperative workloads, allocate resources, and balance loads across computing satellites. The latter host computing accelerators and execute the assigned model components and inference tasks.

\subsubsection{Space Edge Subnetwork}

The space edge subnetwork provides user-facing connectivity and proximity-based computing. It comprises free-flying LEO constellations whose satellites are equipped with large phased-array antennas and ISLs, with selected satellites additionally carrying AI computing payloads.
For communication, the phased-array antennas receive user requests and deliver generated tokens, while the ISLs connect the LEO constellation to gateway satellites in dawn--dusk SSO. Multi-hop routing and satellite handovers thereby sustain request and token exchange between terrestrial users and the SpaceDC subnetwork. For computing, AI-enabled LEO satellites serve nearby users directly. Their proximity to the ground shortens E2E propagation paths, making edge execution suitable for latency-sensitive workloads whose computing and memory demands can be accommodated by a single satellite.

\subsection{Protocol of Space-based Token Service}

\begin{figure*}[t]
    \centering
\subfigure[Constellation Architecture]{
\includegraphics[width=0.34\textwidth]{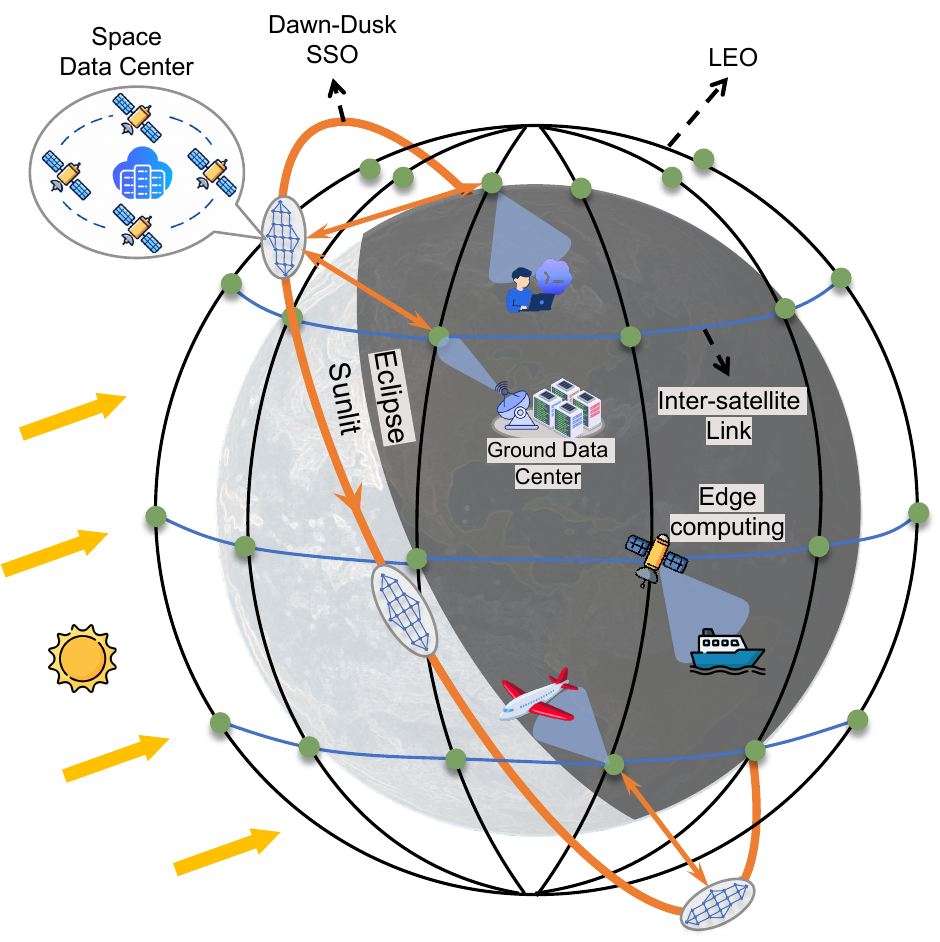}
\label{fig:sdc-constellation}
 }
\subfigure[Protocol of Space-based Token Service]{
\includegraphics[width=0.44\textwidth]{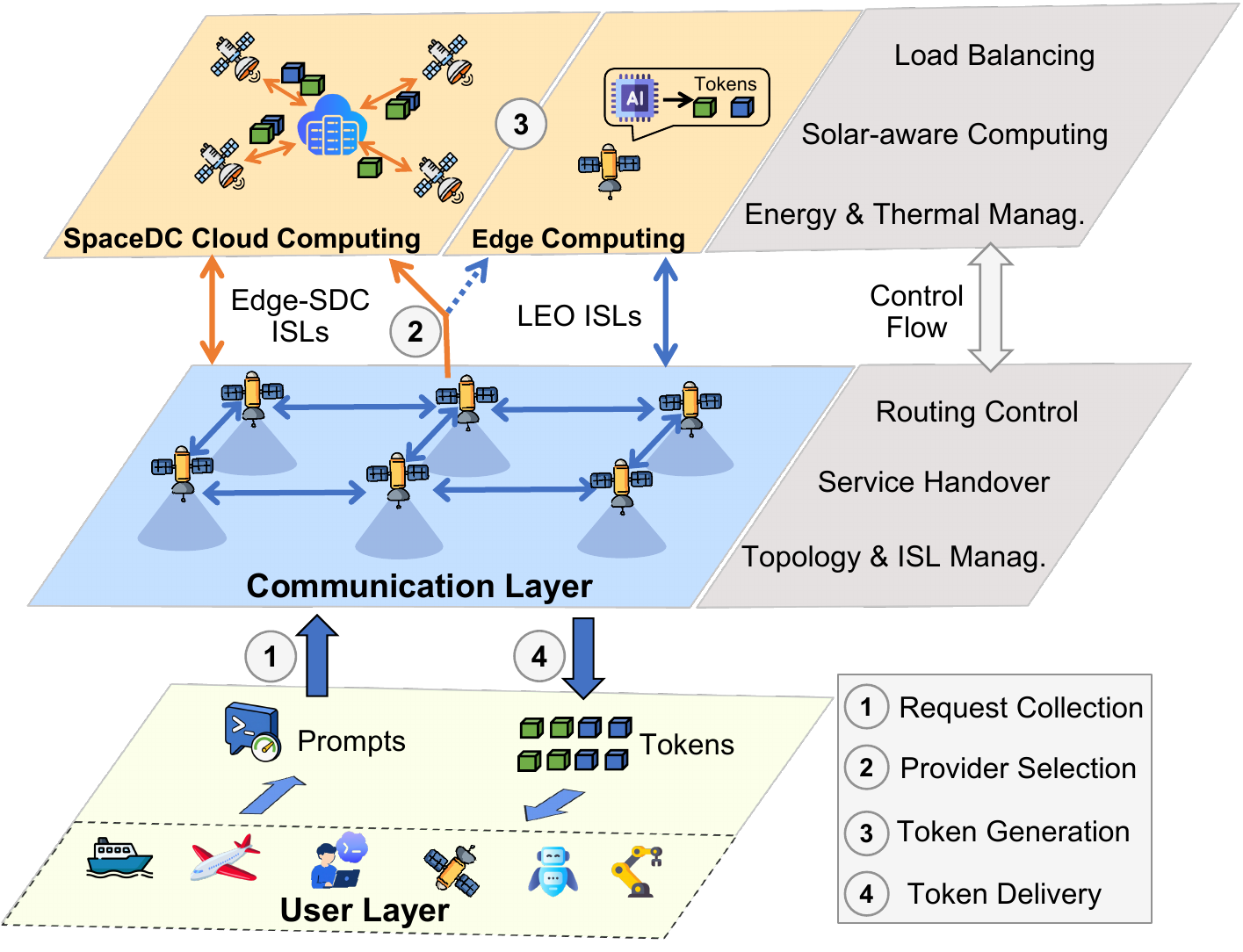}

\label{fig:protocol}
}
   
    \caption{Architecture of Space AI Infrastructure.
      }
  \vspace{-5mm}
\end{figure*}

As shown in Fig.~\ref{fig:protocol}, E2E token services in the proposed architecture operate across two logical planes. The \emph{token plane} carries service data during token generation and delivery, including user prompts, intermediate inference states, and generated tokens. The \emph{control plane} orchestrates these processes through request and token routing, resource allocation, and workload balancing. Together, the two planes realize each token service through four logical stages.

\subsubsection{Token Request Collection}

An access satellite in the space edge subnetwork receives a user prompt and its associated QoS requirements, including the requested AI model, output modality, and token throughput. The prompt enters the token plane, while the service metadata are forwarded to the local control function. The control plane also records the request identifier and the user's serving access satellite for subsequent routing and mobility management.

\subsubsection{QoS-aware Edge--cloud Provider Selection}

The control plane maps the  QoS requirements to computing, memory, and communication demands, together with the corresponding latency constraints. For an MoE-based token service, for example, a higher output-quality target may require activating more experts, thereby increasing the computing workload and inter-satellite traffic during token generation. Based on the available resource and network states, the control plane selects an eligible token provider: either a nearby AI-enabled edge satellite or a SpaceDC. It then reserves the required resources, establishes the forwarding path, and directs the token plane to deliver the prompt to the selected provider.

\subsubsection{Token Generation}

At the selected provider, token generation then proceeds in two phases. The \emph{prefill phase} processes the input-token sequence and constructs the KV cache, whereas the \emph{decoding phase} generates output tokens autoregressively. For a SpaceDC-based service, the gateway satellite orchestrates distributed inference across multiple computing satellites using tensor, pipeline, or other forms of model parallelism. A related example is SpaceMoE, which reduces token-generation latency in LEO networks through topology-aware placement of MoE layers and experts~\cite{Wang2026SpaceMoE}. Intermediate activations and synchronization traffic are confined within the SpaceDC whenever possible, limiting traffic over longer edge--cloud links. For a lightweight edge service that fits within a single satellite, the selected edge satellite performs centralized inference using the allocated resources.

\subsubsection{Token Delivery}

Token delivery is pipelined with autoregressive decoding to reduce both time to first token and inter-token latency. Each newly generated token is detokenized and immediately forwarded through the token plane along the established delivery route. To preserve service continuity, the distributed control plane tracks the association between each active request and its serving access satellite. When user mobility or satellite motion triggers a handover, the control plane updates the delivery route and installs new forwarding entries without restarting token generation. Upon receiving the detokenized output, the serving access satellite streams it to the user over the space--ground downlink.

\begin{figure}[t]
    \centering
                                                \subfigure[CDF of token delivery latency with different fractions of edge computing satellites.]{
                        \includegraphics[width=0.4\textwidth]{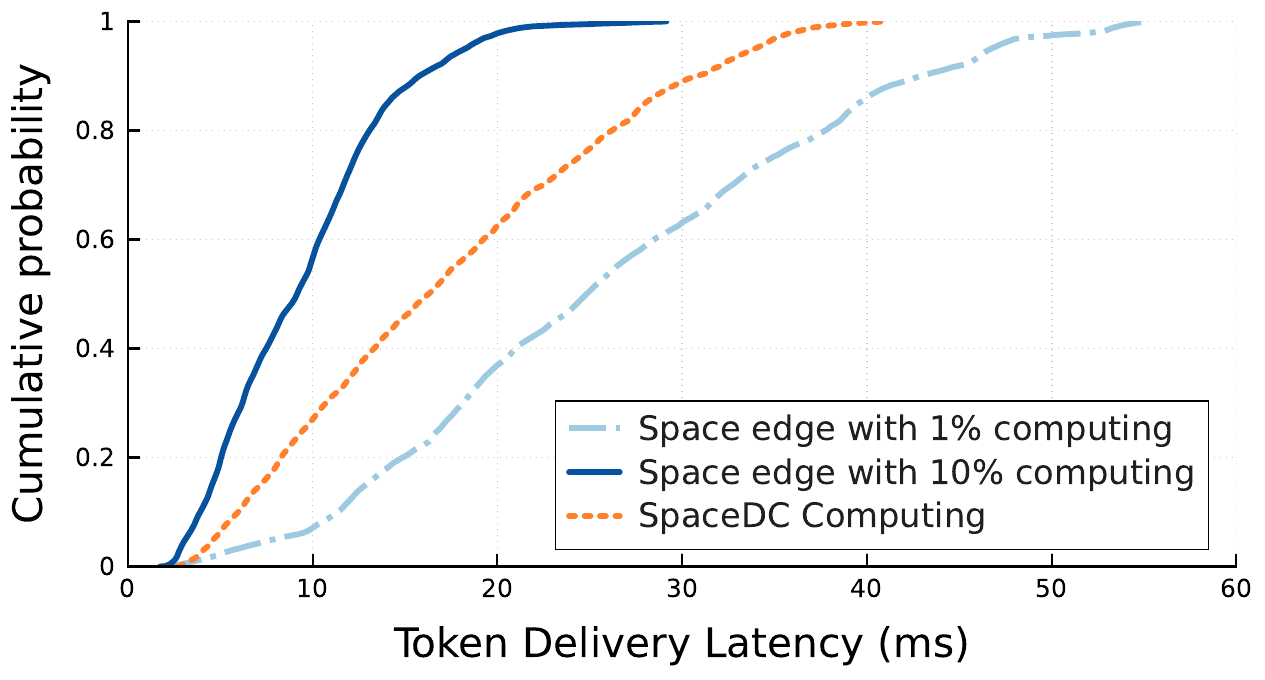}
        \label{fig:selected-token-delivery-latency-cdf}
    }
    \hfill
    \subfigure[Average throughput under a user-demand sweep.]{
                        \includegraphics[width=0.42\textwidth]{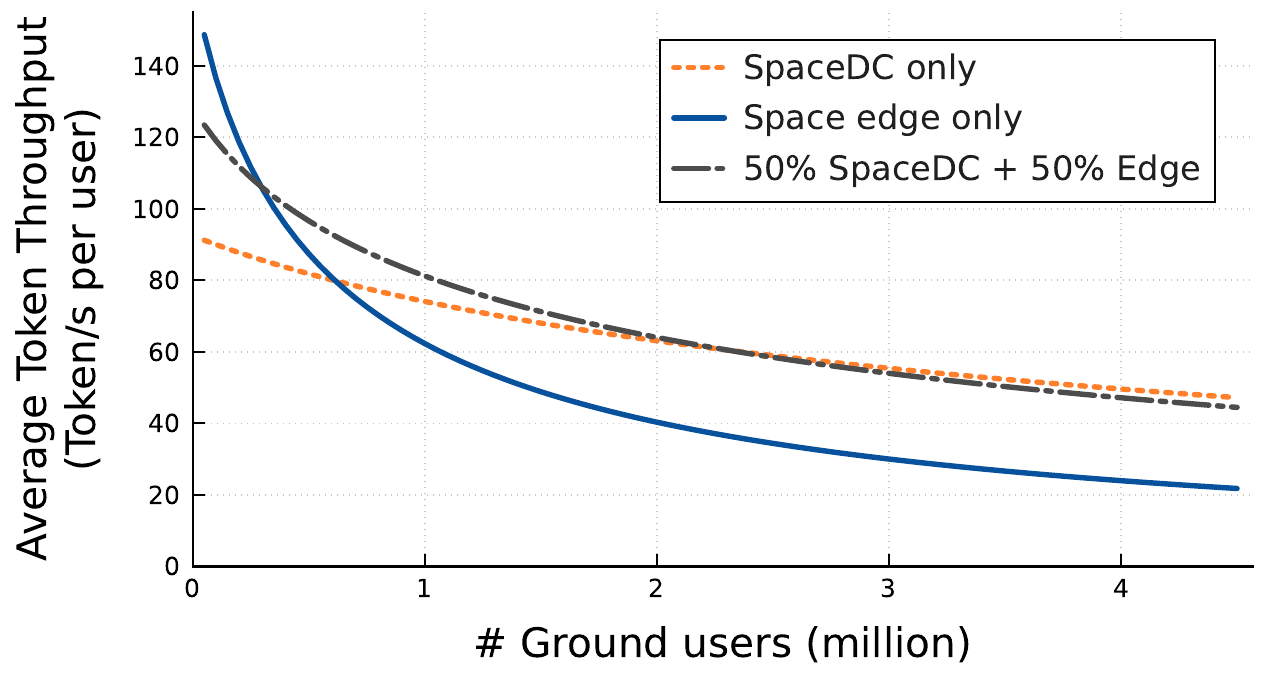}
        \label{fig:selected-user-demand-throughput-curve}
    }

    \vspace{0.5em}
    \subfigure[Throughput density at 10 million users.]{
                                \includegraphics[width=0.42\textwidth]{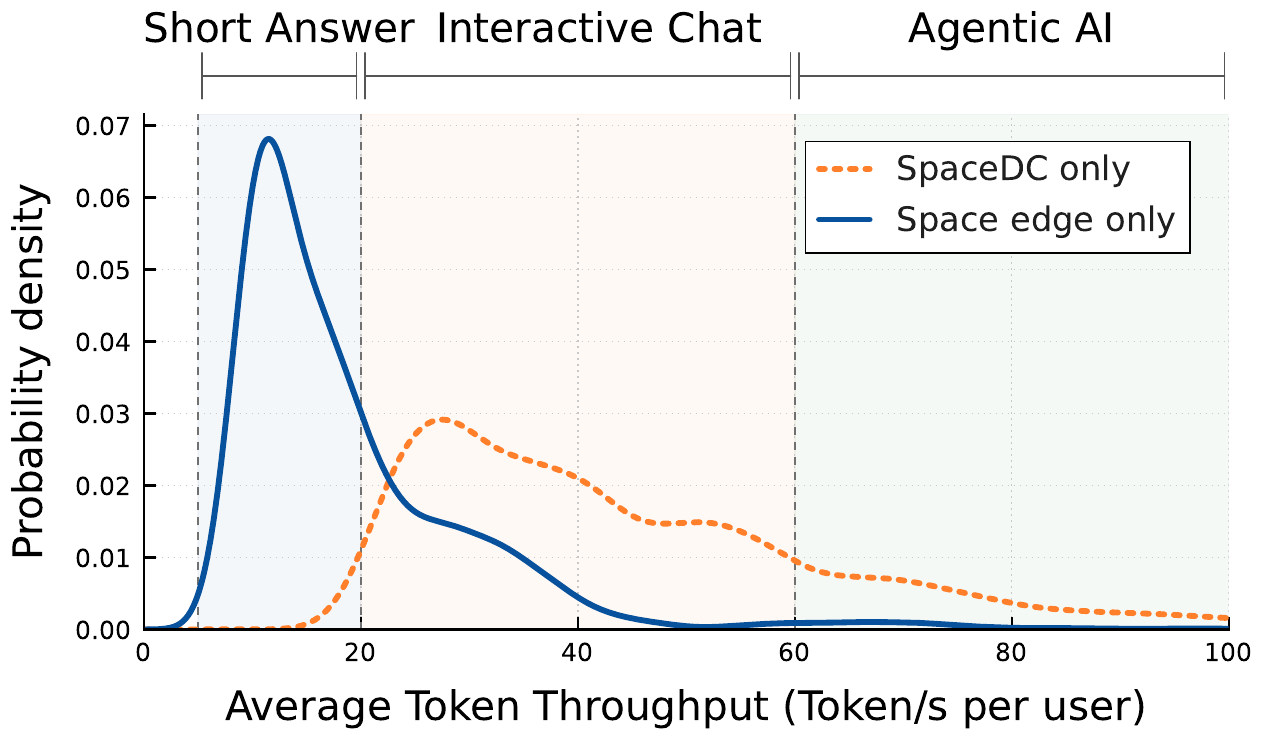}
        \label{fig:selected-token-throughput-pdf-10m}
    }
                                    \caption{Token-service performance \zw{under uniformly distributed demand across Earth's surface}. 
        }
    \label{fig:combined-token-performance-results}
    \vspace{-5mm}
\end{figure}

\section{Case Study: A Comparison between Edge and SpaceDC-based Token Services}
\label{sec:case_study}

\subsection{Experimental Setup}

For the space edge subnetwork, we consider a $640$-satellite polar LEO constellation deployed at an altitude of $500$~km, comprising $32$ orbital planes with $20$ satellites per plane. Each satellite maintains at most four ISLs with adjacent satellites, forming a mesh topology with a linkless counter-rotating seam between the first and last orbital planes. Every active ground user associates with its nearest visible satellite, which serves as the access satellite for its token service. Each access satellite provides a total downlink capacity of $10$~Gbps, shared equally among its associated active users. The SpaceDC subnetwork comprises $20$ dawn--dusk SSO clusters at an altitude of $600$~km, each containing $16$ computing satellites. Each SpaceDC establishes at most four gateway links with visible LEO satellites. The network evolves over one-minute time slots; within each slot, its topology is modeled as a static graph, and tokens are routed along the shortest paths.

For computing, we consider a $70$-billion-parameter dense model with a workload of $140$ billion \emph{floating-point operations} (FLOPs) per generated token. Each SpaceDC computing satellite is assigned a computing throughput equivalent to one NVIDIA GB300 NVL72 rack-scale system, assuming a $100\%$ sunlit fraction. The long-term effective computing throughput of each edge satellite is obtained by scaling this baseline according to its annual sunlit fraction. When a computing satellite concurrently serves multiple users, the latency of one decoding iteration is approximated as the aggregate workload required to generate one token for each served user divided by the satellite's effective computing throughput.

\rev{We evaluate SpaceAI-Infra using token delivery latency and average token throughput. The former is the sum of propagation and transmission delays from the computing satellite to the ground user. The latter is the arithmetic mean of per-user throughputs, each defined as the reciprocal of the E2E token latency, which is the sum of token generation and delivery latencies.}

\subsection{Performance Comparison}

Fig.~\ref{fig:combined-token-performance-results} evaluates the token-serving performance of \sai under uniformly distributed demand across the Earth's surface. Specifically, Fig.~\ref{fig:selected-token-delivery-latency-cdf} plots the \emph{cumulative distribution function} (CDF) of token delivery latency \rev{at $1$ million concurrent users} for different fractions of LEO satellites equipped with edge AI computing, compared with ``SpaceDC Computing.'' At an edge-computing fraction of $10\%$, edge-based service achieves \rev{lower token delivery latency} than its SpaceDC-based counterpart. This gain stems from shorter delivery paths between the edge computing satellites and terrestrial users. Reducing the fraction from $10\%$ to \rev{$1\%$}, however, substantially increases latency, as sparser edge computing satellites require longer routes to deliver tokens to users.

Fig.~\ref{fig:selected-user-demand-throughput-curve} plots the average token throughput against the number of users for the three user-assignment strategies. The throughput decreases with user concurrency as users contend for shared computing and communication resources. At low concurrency, the edge-only case, in which all requests are served by edge satellites, outperforms its SpaceDC-only counterpart. Its throughput, however, degrades more rapidly and eventually falls below that of SpaceDC-based service. This crossover reflects the limited computing capacity and intermittent solar availability of edge satellites, whose widely separated geometry also prevents efficient resource pooling. By contrast, SpaceDCs sustain relatively stable throughput by pooling formation-flying computing nodes under near-continuous solar exposure. The equal-split case exceeds both single-tier cases at intermediate concurrency, but falls below SpaceDC-only service at the highest concurrency.

Fig.~\ref{fig:selected-token-throughput-pdf-10m} plots the \emph{probability density function} (PDF) of user token throughput under \rev{$5$} million concurrent users. The SpaceDC-only case provides most users with sufficient throughput for interactive chat\rev{, with a smaller fraction supporting} agentic AI services. Edge-only operation is largely limited to short-answer services, with \rev{approximately $32\%$ of users} attaining the throughput required for interactive chat. These results highlight the essential role of SpaceDCs in supporting Space Cloud AI under massive user concurrency.

\begin{figure*}[t]
    \centering
                                                                                \includegraphics[width=0.75\textwidth]{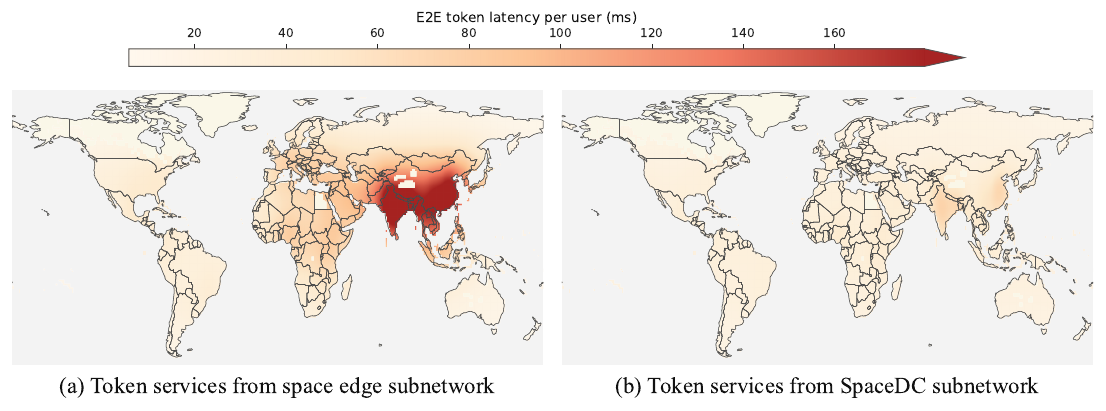}
    \caption{Yearly-average E2E token latency per user under population-weighted token demand with the polar LEO access constellation. 
            }
    \label{fig:selected-world-population-e2e-latency-heatmap-yearly-average-polar}
    \vspace{-5mm}
\end{figure*}

Fig.~\ref{fig:selected-world-population-e2e-latency-heatmap-yearly-average-polar} maps the geographical distribution of \rev{yearly-average} E2E token latency for services from space edge (left) and SpaceDCs (right), using a shared color scale, when \rev{demand from $2$ million concurrent users} follows the Gridded Population of the World population-count data~\cite{CIESIN2018GPWv4}. Under space edge service, densely populated regions, particularly East and South Asia, experience E2E token latencies exceeding $100$~ms per token. Concentrated demand in these regions saturates the limited computing capacity of the overflying edge AI satellites. By pooling substantially greater computing capacity, SpaceDCs alleviate these regional bottlenecks.

\section{Concluding Remarks}
\label{sec:conclusion}
We have proposed an edge–cloud-integrated architecture for SpaceAI-Infra. 
Its key feature lies in the confluence of communication-centric and computing-centric satellite constellations for token services in space. 
In particular, LEO-based free-flying constellations exploit their spatial distribution over the Earth to provide ubiquitous user access and global token delivery, while their long inter-satellite distances limit the tightly coupled computing required for data-center-scale token generation. This limitation is addressed by formation-flying satellites operating in proximity along dawn–dusk sun-synchronous orbits, forming promising SpaceDCs for Space Cloud AI. Integrated with the space edge, these SpaceDCs are capable of delivering generated tokens to users anywhere on Earth.
The proposed architecture raises several research directions.
\begin{itemize}
    \item \textbf{Computing-efficient Topology of SpaceDCs:} One emerging direction is the largely unexplored network topology of formation-flying satellite constellations. Its design requires a fusion of orbital dynamics and computing-efficient interconnect structures. For example, terrestrial butterfly or hierarchical interconnects among GPU racks may not readily satisfy satellite visibility and hardware constraints. This calls for new SpaceDC topologies and associated ISL control tailored to parallel computing. Such ISL control should support data-center-scale token exchange while remaining robust to radiation-induced link and node failures.
    
    \item \textbf{Space Edge-cloud Cooperative Token Generation:} 
    The cooperation is enabled by distributing foundation models across edge satellites and SpaceDCs for cooperative inference.
Its implementation faces unique challenges different from terrestrial edge–cloud systems, such as asymmetric solar availability, limited ISL ports, and time-varying edge–cloud visibility. 
Addressing these requires rethinking the LLM inference workflow and exploiting inference-level degrees of freedom, such as model placement and token routing. Moreover, as cloud computing scales, the communication bottleneck may shift from space–ground access to edge–cloud ISLs. This shift calls for revisiting LEO constellation coverage: the space edge must cover not only terrestrial users, but also dawn–dusk SSO data centers.
\end{itemize}

\bibliographystyle{IEEEtran}
\bibliography{Ref.bib}

\section*{Biographies}
\begingroup
\small

\noindent\textbf{Zhanwei Wang} is a research assistant professor with the Department of Electrical and Computer Engineering, The University of Hong Kong, Hong Kong. His research interests include wireless communications, edge intelligence, and space AI.
\par\medskip

\noindent\textbf{Jianhao Huang} is a postdoctoral fellow with the Department of Electrical and Computer Engineering, The University of Hong Kong, Hong Kong. His research interests include compressed sensing, semantic communications, and deep learning.
\par\medskip

\noindent\textbf{Min Sheng} [Fellow, IEEE] is a full professor and the director of the State Key Laboratory of Integrated Service Networks, Xidian University, Xi'an, China. Her research interests include mobile ad hoc networks, 5G mobile communication systems, and satellite communication networks.
\par\medskip

\noindent\textbf{Kaibin Huang} [Fellow, IEEE] is a professor with the Department of Electrical and Computer Engineering, The University of Hong Kong, Hong Kong. His research interests include mobile edge computing, edge AI, and 6G systems.
\par
\endgroup

\end{document}